\documentclass[12pt]{iopart}

\usepackage{iopams}

\newcommand{\abs}[1]{\ensuremath{\left|#1\right|}}

\newcommand{\expv}[1]{\ensuremath{\langle#1\rangle}}

\newcommand{\kB}{k_{\rm B}}

\newcommand{\gmin}{\gamma_{\rm min}}
\newcommand{\gmax}{\gamma_{\rm max}}

\newcommand{\eqref}[1]{(\ref{#1})}

\begin{document}
\title[]{Decoherence from ensembles of two-level fluctuators}

\author{Josef Schriefl$^1$, Yuriy Makhlin$^2$, Alexander Shnirman$^1$, and Gerd Sch\"on$^1$}

\address{$^1$ Institut f\"{u}r Theoretische Festk\"{o}rperphysik,
Universit\"{a}t Karlsruhe, D-76128 Karlsruhe, Germany}
\address{$^2$ Landau Institute for Theoretical Physics, Kosygin st. 2, 119
334 Moscow, Russia}

\begin{abstract}
$1/f$ noise, the major source of dephasing in Josephson qubits, may
be produced by an ensemble of two-level systems. Depending on the
statistical properties of their distribution, the noise distribution
can be Gaussian or non-Gaussian. The latter situation is realized, for
instance, when the distribution of coupling strengths has a
slowly decaying power-law tail. In this regime questions of
self-averaging  and sample-to-sample fluctuations become crucial. We
study the dephasing process for a class of distribution functions
and analyze the self-averaging properties of the results.
\end{abstract}
\pacs{}
\submitto{New Journal of Physics}
\maketitle
\normalsize

\section{Introduction}

In Josephson qubits dephasing is dominated by low-frequency noise,
often with a $1/f$ power spectrum, due to fluctuations of background charges,
magnetic fluxes,
or critical currents~\cite{Nakamura_Echo,Saclay_Science,Clarke}.
While irrelevant for the relaxation process with time scale $T_1$, low-frequency noise
dominates the dephasing time $T_2^*$. Standard NMR echo techniques
allow one to reduce dephasing by rendering the low frequency spectrum
ineffective~\cite{Nakamura_Echo}. Operation at optimal
bias points, chosen such that the linear longitudinal coupling of the qubit to
the $1/f$ noise source vanishes, proved to be very successful in
increasing the dephasing time~\cite{Saclay_Science}. Further
progress in this direction may require an improved understanding of the
mechanisms causing $1/f$ noise and of its statistical properties. It
was realized recently that qubits themselves can be used to study
the noise properties of their
environment~\cite{Aguado,SchoelkopfNoiseDetector}, and an
interesting relation between the low-frequency $1/f$ and the
high-frequency charge noise was observed~\cite{Astafiev}. An
extensive study of dephasing due to both charge and flux noise
was undertaken in Ref.~\cite{Ithier2005}.

Still, many questions remain open. If the number of fluctuators
contributing to the $1/f$ noise is large, one could expect Gaussian
statistics~\cite{Nakamura_Echo,Our_PhysScripta}.
In Ref.~\cite{Paladino_PRL02} and following work the role of individual,
strongly coupled fluctuators was emphasized, and it was suggested~\cite{Galperin}
that even  ensembles of many fluctuators may produce strong non-Gaussian
effects, emerging as a result of rare configurations in which dephasing is
dominated by a small number of very strongly coupled fluctuators.
As far as we can judge, the decay laws observed in
Ref.~\cite{Ithier2005} cannot be fully explained by either of these theories.

As the experiments are performed on individual systems with a
particular configuration of the fluctuators, it is important to
understand whether the predicted decay laws are self-averaging or
have strong sample-to-sample fluctuations. Here we will analyze a
class of distribution functions for the coupling strengths of the
fluctuators. We determine the ensemble-averaged decay laws (extending the
results of Ref.~\cite{Galperin}) and analyze which of them are
self-averaging. We study both dephasing due to linear longitudinal
coupling and dephasing at the optimal point where the coupling is
quadratic.

\section{$1/f$ noise from two-level fluctuators (TLF)}

$1/f$ noise is often attributed to a collection of bistable systems,
switching randomly between two states~\cite{Bernamont}.
On one hand, such a model provides a natural explanation of $1/f$ noise.
On the other hand, in many samples distinct two-level fluctuators
(TLF's) were detected. In metals this switching causes conductance
fluctuations~\cite{Ludviksson,Kogan} and, consequently, $1/f$ noise
of the transport current. In Josephson junctions it causes the critical
current to fluctuate~\cite{Kozub,Galperin_Gurevich_Kozub}. More generally, 
spin bath environments were analyzed in Ref.~\cite{Stamp}.
In charge qubits the TLF's contribute to the fluctuations of
the gate charge controlling the qubit. The TLF's are characterized
by their coupling strengths to the qubit, $v_n$, which may vary depending on the
location of the respective TLF. The fluctuating quantity that
couples to the qubit, $X(t)$, contains contributions from all TLF's:
\begin{equation}
\label{eq:X_sum}
X(t)=\sum_{n}v_{n}\sigma_{n,z}(t)
\end{equation}
Each fluctuator switches randomly between two positions, denoted by
$\sigma_{n,z}=\pm1$, with rate $\gamma_{n}$ (for simplicity, we
assume equal rates in both directions for the relevant TLF's) and
thus contributes to the noise power $S_{X}(\omega)=\sum_{n}S_{n}(\omega)$ with
\begin{equation}
S_{n}(\omega)=\frac{2\gamma_{n}v_{n}^{2}}{\omega^{2}+\gamma_{n}^{2}}\,.
\end{equation}

A set of TLF's produces $1/f$ noise when the switching rate $\gamma$
depends exponentially on a physical quantity, $l$, with
a smooth distribution. For instance,  $\gamma\propto
e^{-l/l_0}$, with $l$ distributed uniformly over a range much wider
than $l_0$,
translates in a log-uniform distribution of the switching rates,
with probability density $P(\gamma)\propto 1/\gamma$ in the
corresponding exponentially wide range $\gamma_{\rm
min}\ll\gamma\ll\gamma_{\rm max}$. In this range the total noise
power thus scales as
\begin{equation}
  \label{eq:Sgamma}
 S_{X}(\omega) \propto \int \frac{d\gamma}{\gamma} \,
\frac{2v^2\gamma}{\gamma^2 + \omega^2} \,\propto\,
\frac{v^2}{\abs{\omega}}\ .
\end{equation}

An example is a particle trapped in a double-well potential, whose
tunneling rate through the potential barrier depends exponentially
on both the height and the width of the barrier, leading to $1/f$
noise. Another example is  thermally activated tunneling with
rate $\gamma_0 e^{-E/\kB T}$, where $E$ denotes an activation
energy. In this way the $1/f$ power spectrum observed in metals can be
attributed to a broad (much wider than $\kB T$) distribution of activation
energies~\cite{DuttaHorn}.

\section{Distribution of coupling strengths, self-averaging}

The analysis of decoherence in the presence of many fluctuators
requires the study of probability distributions of coupling strengths
and switching rates. In
each particular sample one deals with specific fluctuators,
i.e., with a realization  of the set of parameters $v$ and $\gamma$, drawn
from this distribution. One should distinguish between quantities
averaged over a statistical ensemble of samples and the results for
a specific sample. This difference is essential, if the quantity
under consideration is not self-averaging, i.e., if it has considerable
sample-to-sample fluctuations. Such a situation arises if a quantity
is dominated by contributions from a small number of TLF's.

In Ref.~\cite{Galperin}, a continuous distribution of the parameters $v_{n}$ and
$\gamma_{n}$ was considered, with a long tail of the distribution
of coupling strengths $v_{n}$, such that rare configurations
with very large $v_n$ dominate certain ensemble properties. It
arises, e.g., from a uniform spatial distribution of fluctuators on a
$d$-dimensional surface and a power-law TLF-qubit coupling~\cite{Galperin}, $v(r) \propto
1/r^b$. This results in a distribution of coupling strengths
$P(v) \propto 1/v^{1+d/b}$. The joint distribution $P(v,\gamma)$,
defined in the domain
$[v_{min},\infty]\times\lbrack\gamma_{min},\gamma _{max}]$ and
normalized to describe $N$ fluctuators is thus
\begin{equation}
  \label{eq:Pvgamma}
  P(v,\gamma) = \frac{c}{\gamma}\,\frac{\mu\eta^\mu}{v^{1+\mu}} \ .
\end{equation}
Here $\mu = d/b>0$, $c=1/\ln(\gmax/\gmin)$, and $\eta
 = v_{\rm min}  N^{1/\mu}$.
One can also allow for fluctuations of $N$, but this does not
change the results significantly.

We consider a $d$-dimensional volume of typical size $r_{\rm max}$
around the qubit containing a uniform distribution of TLF's. The
typical distance between the strongest (closest) fluctuator and the qubit thus
scales as $r_{\rm min} \sim (V/N)^{1/d} \sim r_{\rm max}/N^{1/d}$.
On the other hand, since the coupling strength was assumed to decay
as $v(r) \propto 1/r^b$ the relation between the strongest and
weakest coupling strength is given by $v_{\rm max}/v_{\rm min} =
(r_{\rm max}/r_{\rm min})^b$. Combining both results we find that
the typical maximal coupling strength scales as $v_{\rm max}^{\rm
typ} \sim v_{\rm min} N^{1/\mu}$. This does not exclude the existence of fluctuators
with $v\gg v_{\rm max}^{\rm typ}$ in certain
realizations, as the long tail of the distribution function suggests.

As examples of averaging over the distribution of coupling
strengths and switching rates we calculate the noise produced by
the ensemble of fluctuators,
\begin{equation}
  \label{eq:Sintegr}
 S_{X}(\omega) = \int d v\,d\gamma  \,P(v,\gamma) \,
\frac{2v^2\gamma}{\gamma^2 + \omega^2}
\end{equation}
distinguishing two cases: In one case, for $\mu<2$ the integral over $v$
diverges at the upper limit. Hence the noise is dominated by the
strongest fluctuator(s). Thus the result is sensitive to the properties
of one or a few fluctuators and is therefore not self-averaging.
Estimates below are based on cutting the integral
at $v=\eta=v_{\rm max}^{\rm typ}$ but one has to remember that for a
comparison with experiments
averaging (including the averaging over $\gamma$) makes little
sense, since only a few TLF's contribute.

In contrast, for $\mu>2$ the integral is dominated by fluctuators
with $v < \eta$. The weak fluctuators are most important, and due
to their large number the noise is given by a sum of many comparable
independent contributions. Consequently, the result is self-averaging, i.e., in
different samples or runs of the experiment with $\mu > 2$ one should
observe the same noise amplitude.

We now summarize the typical/average results for the noise, retaining only the
leading contributions
\begin{equation}
  \label{eq:SmanyBF2}
  S_X(\omega) = \frac{2\pi A}{\abs\omega} \quad {\rm with} \quad
A = \left\{
\begin{array}{l@{\quad : \quad}l}
c\;\frac{\mu}{2-\mu} \,
\eta^2 & \mu < 2\vspace*{0.25cm} \ \ (\mbox{typically})\\
c\; N \langle v^2 \rangle & \mu > 2\\
\end{array}\right.\,.
\end{equation}
For $\mu>2$ we defined the average coupling strength of the TLFs,
$\langle v^2 \rangle = \frac{1}{N} \int dv P(v) v^2$. Note that
\eqref{eq:SmanyBF2} is only valid for frequencies $\gmin \ll
\abs\omega \ll \gmax$. At lower frequencies, $\abs\omega <\gmin$,
$S_X(\omega)$ tends to a constant, whereas at higher frequencies,
$\abs\omega > \gmax$,  $S_X(\omega)$ crosses over to a faster power
law decay $\propto 1/\omega^2$.

\section{Longitudinal and transverse noise coupling}

We consider a qubit controlled (for simplicity) by a single parameter
$\lambda$ and Hamiltonian
\begin{equation}
\label{Hspin}
{H}_{qb}=-\frac{1}{2}\vec{H}_{0}(\lambda)\vec{\sigma}\, .
\end{equation}
After an initial preparation in a coherent superposition of the
qubit's eigenstates, the effective spin precesses under the
influence of the static field $\vec{H}_{0}$, set by the control
parameter $\lambda_{0}$. Coupling to the environment disturbs this
evolution, leading to decoherence. In many cases the effect of the
environment can be modeled by classical and quantum fluctuations of
$\lambda(t) = \lambda_0 +X(t)$, where ${X(t)}$ fluctuates. For
instance, in a charge qubit electromagnetic fluctuations of the
control circuit as well as the background charge noise influence the
gate voltage which controls the qubit.

To proceed we expand the Hamiltonian ${H}_{qb}$ to second order in
the perturbation ${X}$,
\begin{equation}
{H}_{qb}=-\frac{1}{2}\left[  \,\vec{H}_{0}(\lambda_{0})+\frac{\partial
\vec{H}_{0}}{\partial\lambda}\,{X}+\frac{\partial^{2}\vec{H}_{0}%
}{\partial\lambda^{2}}\,\frac{{X}^{2}}{2}+...\right]  \,\vec
{\sigma}\,.
\end{equation}
Introducing the notations $\vec{D}_{\lambda}\equiv(1/\hbar)\,\partial\vec
{H}_{0}/\partial\lambda$ and $\vec{D}_{\lambda2}\equiv(1/\hbar)\,\partial
^{2}\vec{H}_{0}/\partial\lambda^{2}$, we find in the eigenbasis of $\vec
{H}_{0}(\lambda_{0})\,{\vec\sigma}$:
\begin{equation}
{H}_{qb}=-\frac{1}{2}\hbar\left(  \omega_{01}{\sigma}%
_{z}+\delta\omega_{z}{\sigma}_{z}+\delta\omega_{\perp}{\sigma
}_{\perp}\right) \label{Eq:Heigen}%
\end{equation}
where $\hbar\omega_{01}\equiv|\vec{H}_{0}(\lambda_{0})|$,
$\delta\omega _{z}\equiv
D_{\lambda,z}{X}+D_{\lambda2,z}\,{X}^{2}/2+... $, and
$\delta\omega_{\perp}\equiv D_{\lambda,\perp}{X}+...$. Here
$\sigma_{\perp}$ denotes the transverse spin components (i.e., $\sigma_{x}$ or
$\sigma_{y}$). The coefficients $D$ are related to the derivatives
of $\omega_{01}(\lambda)$:
\begin{equation}
\frac{\partial\omega_{01}}{\partial\lambda}=D_{\lambda,z}\ ,
\end{equation}
and
\begin{equation}
\frac{\partial^{2}\omega_{01}}{\partial\lambda^{2}}=%
D_{\lambda2,z}+\frac{\vec D_{\lambda,\perp}^{2}}{\hbar\omega_{01}}\,.%
\end{equation}
Thus, in general, the coupling of noise to the qubit contains both
transverse ($\delta\omega_{z}$) and longitudinal
($\delta\omega_{\perp}$) parts, and both may have linear as well as
higher order (e.g., quadratic) contributions.

\section{Bloch-Redfield theory}

For weak, short-correlated noise the dynamics of the two-level
systems (spins, qubits) can be summarized by Bloch's
equations~\cite{Bloch_Derivation,Redfield_Derivation} in terms
of two rates: the longitudinal relaxation (depolarization) rate
$\Gamma_{1}=T_{1}^{-1}$, and the transverse relaxation (dephasing)
rate $\Gamma_{2}=T_{2}^{-1}$. Evaluated perturbatively, using the
golden rule, the rates are given by
\begin{equation}
\Gamma_{1}=\frac{1}{2}
S_{\delta\omega_{\perp}}(\omega=\omega_{01})=\frac{1}{2}
D_{\lambda,\perp}^{2}\,S_{X}(\omega=\omega_{01})%
\end{equation}
and
\begin{equation}
\Gamma_{2}=\frac{1}{2}\Gamma_{1}+\Gamma_{\varphi}\ ,\label{Gamma2}%
\end{equation}
where
\begin{equation}
\Gamma_{\varphi}=\frac{1}{2} S_{\delta\omega_{z}}(\omega=0)=\frac{1}{2} D_{\lambda,z}%
^{2}\,S_{X}(\omega=0)\ .\label{white}%
\end{equation}
The dephasing process (\ref{Gamma2}) is a combination of
depolarization effects ($\Gamma_{1}$) and of the so called `pure'
dephasing, characterized by the rate $\Gamma_{\varphi}={T_2^{*}}^{-1}$. The
pure dephasing is usually associated with the inhomogeneous level
broadening in ensembles of spins, but occurs also for a single
spin due to the longitudinal low-frequency noise.

\section{Pure dephasing for Gaussian noise, $\mu >2$}

If there are sufficiently many fluctuators in the environment, the
central limit theorem (CLT) applies, and the noise is Gaussian. More
specifically, since the central limit theorem applies to a large
collection of equally distributed random quantities, one needs to
have a large number of TLF's of each (relevant) ``kind'' (i.e., for
each pair $v$, $\gamma$). This implies a regular distribution of coupling
strengths, so that the relevant physical quantities are not
dominated by a few TLF's at a boundary of the distribution.
In particular, the distribution (\ref{eq:Pvgamma})
gives rise to Gaussian noise if $\mu >2$. We will discuss now the pure
dephasing derived from such Gaussian noise. The
random phase accumulated at time $t$,
\[
\Delta\phi=D_{\lambda,z}\int\limits_{0}^{t}dt^{\prime}{X}(t^{\prime
}) \,
\]
is then also Gaussian-distributed. Hence the decay law, due to
longitudinal noise (coupling to $\sigma_z$) in a free
induction decay (Ramsey signal) is given by
$f_{R}(t)=\langle\exp(i\Delta\phi)\rangle
=\exp(-(1/2)\langle\Delta\phi^{2}\rangle)$.
Averaging here is over the different trajectories of
$X(t)$ in repeated runs of the dephasing experiment.
We obtain
\begin{equation}
f_{R}(t)=\mathop{\rm exp}  \nolimits\left[  -\frac{t^{2}}{2}\,D_{\lambda
,z}^{2}\int_{-\infty}^{+\infty}\frac{d\omega}{2\pi} S_{X}(\omega)\,\mathop{\rm sinc}
\nolimits^{2}\frac{\omega t}{2}\right]  \,,\label{ramsey1rst}%
\end{equation}
where $\mathop{\rm sinc}  \nolimits x\equiv\sin x/x$. If most of the
noise power is concentrated at frequencies $\omega \ll 1/t$ (static noise), then
one can approximate $\mathop{\rm sinc}\frac{\omega t}{2} \approx 1$
and obtain
\begin{equation}
f_{R}^{\mathrm{stat}}(t)=\mathop{\rm exp}  \nolimits\left[  -\frac{t^{2}}%
{2}\,D_{\lambda,z}^{2}\sigma_{X}^{2}\right]  \,,\label{ramsey_static}%
\end{equation}
where $\sigma_{X}^{2}=\int_{-\infty}^{+\infty}(d\omega/2\pi) S_{\lambda}%
(\omega)$ is the dispersion of ${X}$.
In general, for static noise with (not necessarily Gaussian) distribution function $P(X)$
the Ramsey decay is given by
\begin{equation}
f_{R}^{\mathrm{stat}}(t)=\int d({X})P({X}
)\,e^{iD_{\lambda,z}\,{X}\,t}\,,\label{ramsey_static_PX}%
\end{equation}
i.e., by the Fourier transform of $P({X})$. Static noise corresponds
to a situation when $X$ is constant during each run of the
experiment but fluctuates between different runs.

In an echo experiment, the phase acquired is the difference between the two
free evolution periods:
\begin{equation}
\Delta\phi_{E}=-\Delta\phi_{1}+\Delta\phi_{2}=-D_{\lambda,z}\int
\limits_{0}^{t/2}dt^{\prime}{X}(t^{\prime})+D_{\lambda,z}%
\int\limits_{t/2}^{t}dt^{\prime}{X}(t^{\prime})\,,
\end{equation}
which after averaging over the trajectories of $X(t)$ gives
\begin{equation}
f_{E}(t)=\mathop{\rm exp}  \nolimits\left[  -\frac{t^{2}}{2}\,D_{\lambda
,z}^{2}\int_{-\infty}^{+\infty}\frac{d\omega}{2\pi} S_{X}(\omega)\sin^{2}\frac{\omega
t}{4}\mathop{\rm sinc}  \nolimits^{2}\frac{\omega t}{4}\right]
.\label{echo1rst}%
\end{equation}

\textit{$1/f$ spectrum:} Here and below  we assume that the $1/f$ law extends over a wide
range of frequencies, limited by infrared  and ultraviolet
cut-offs,
\begin{equation}
S_{\lambda}(\omega) = \frac{2\pi A}{|\omega|}= \frac{A}{|\nu|},
\quad{\rm
for}\quad\omega_{\mathrm{ir}}\ll|\omega|\ll\omega_{\mathrm{c}}\,.
\end{equation}
The infra-red cutoff $\omega_{\mathrm{ir}}$ is usually determined by
the measurement protocol, as discussed further below. The decay
rates typically depend only logarithmically on
$\omega_{\mathrm{ir}}$, and details of the noise
power below $\omega_{\mathrm{ir}}$ are irrelevant to logarithmic
accuracy. For most of our analysis, the same applies to
the ultra-violet cut-off $\omega_{c}$. However, for some specific
questions considered below, frequency integrals may be dominated by
$\omega\approx \omega_{\mathrm{c}}$, and thus the detailed behavior
near and above $\omega _{\mathrm{c}}$ (i.e. the ``shape'' of the cut-off) is
relevant. We will refer to an abrupt suppression above
$\omega_{\mathrm{c}}$ ($S(\omega
)\propto\theta(\omega_{\mathrm{c}}-|\omega|)$) as a `sharp cut-off',
and to a crossover at $\omega\sim\omega_{\mathrm{c}}$ to a faster
decay $1/\omega\to1/\omega^{2}$ (motivated by modeling of the noise
via a set of bistable fluctuators, see below), as a `soft cut-off'.

For $1/f$ noise, at times $t\ll1/\omega_{ir}$, the free induction (Ramsey)
decay is dominated by the frequencies $\omega<1/t$, i.e., by the quasistatic
contribution~\cite{ACthesis}, and (\ref{ramsey1rst}) reduces to
\begin{equation}
f_{R}(t)=\mathop{\rm exp}  \nolimits\left[  -t^{2}\,D_{\lambda,z}%
^{2}\,A\,\left(  \ln\frac{1}{\omega_{ir}t}+O(1)\right)  \right]
.\label{ramsey1f}%
\end{equation}
Here the logarithmically large part of the exponent
originates from a static contribution of frequencies $\omega < 1/t$.
Indeed it can be obtained from Eq.~(\ref{ramsey_static}) with
$\sigma_{X}^{2}=2 \int_{\omega_{\rm ir}}^{1/t}(d\omega/2\pi) S_{X}%
(\omega)=
\,A \, \ln(1/\omega_{ir}t)$. This contribution dominates the decay of $f_{R}(t)$.

For the echo decay we obtain
\begin{equation}
f_{E}(t)=\mathop{\rm exp}  \nolimits\left[  -t^{2}\,D_{\lambda,z}%
^{2}\,A\cdot\ln2\right]  \ .\label{ramsey1fecho}%
\end{equation}
The echo method thus increases the decay time only by a logarithmic
factor. This low efficiency of the echo has its origin in the
high-frequency tail of the $1/f$ noise, which, as we note,
influences the results strongly. For $1/f$ noise with a
low cut-off $\omega_{\mathrm{c}}$ the integral in
Eq.~(\ref{echo1rst}) over the interval
$\omega\lesssim\omega_{\mathrm{c}}$ is dominated by the upper limit.
For instance, in the
case of a sharp cutoff, i.e., $S=(A/|\omega|)
\theta(\omega_{c}-\omega)$, we obtain
\begin{equation}
f_{E}(t)=\mathop{\rm exp}  \nolimits\left(  -\frac{1}{32}\,D_{\lambda,z}%
^{2}\,A\,\omega_{c}^{2}\,t^{4}\right)  \,.\label{echo_static}%
\end{equation}
On the other hand, for a soft cut-off, which we expect when the noise is
produced by a collection of bistable fluctuators with Lorentzian
spectrum, the integral in
Eq.~(\ref{echo1rst}) is dominated by frequencies $\omega_{\mathrm{c}}%
<\omega<1/t$, and we find $\ln f_{E}(t)\propto D_{\lambda,z}^{2}%
\,A\,\omega_{c}\,t^{3}$. In either case, one finds that the decay is slower
 by a factor
$\sim(\omega_{\mathrm{c}}t)^{2}$ or $\omega_{\mathrm{c}}t$,
respectively, than for $1/f$ noise with a high cutoff,
$\omega_{\mathrm{c}}> D_{\lambda,z} A^{1/2}$.

\section{Individual fluctuators}

We consider a single fluctuator coupled longitudinally to the
qubit, whose contribution to the
level splitting, $v_n(t)\equiv v_n \sigma_{n,z}(t)$, switches between $\pm v_n$.
For this case the free induction (Ramsey) and echo decays have been
evaluated in Refs.~\cite{Paladino_PRL02,Galperin}. In the limit of high
effective temperature, i.e., when the transition rates in both
directions are equal, the decay functions, obtained by averaging over
the switching history of $\sigma_{n,z}(t)$, are given by
\begin{equation}
f_{R,n}(t)=e^{-\gamma_{n}t}\left(  \cos\mu_{n}t+\frac{\gamma_{n}}{\mu_{n}%
}\sin\mu_{n}t\right)  \ ,\label{SingleFluct_R}
\end{equation}
and
\begin{equation}
f_{E,n}(t)=e^{-\gamma_{n}t}\left(
1+\frac{\gamma_{n}}{\mu_{n}}\sin\mu
_{n}t+\frac{\gamma_{n}^{2}}{\mu_{n}^{2}}(1-\cos\mu_{n}t)\right)
\ ,\label{SingleFluct_E}%
\end{equation}
where $\mu_{n}\equiv\sqrt{(Dv_{n})^{2}-\gamma_{n}^{2}}$ and $D\equiv
D_{\lambda,z}$.  In order to derive these expressions we introduce
the averaged phase factors $\chi_\pm(t)= \langle \exp(i\int\limits^t dt'\, D\,
v_n(t')) \rangle$,  averaged over the
switching histories ending at $v_n(t)=+v_n$ or $-v_n$, respectively.
Their dynamics is governed by the rate equations
\begin{eqnarray}
\nonumber
\dot\chi_+ = \phantom{-}i Dv_n \; \chi_+ - \gamma_n \chi_+ + \gamma_n
\chi_- \ ,\\
\dot\chi_- = -i Dv_n \; \chi_- - \gamma_n \chi_- + \gamma_n \chi_+
\,.\label{Eq:chi-}
\end{eqnarray}
The solution for $f_{R,n}(t)=\chi_+(t)+\chi_-(t)$ is obtained
by solving the coupled equations for the initial conditions
$\chi_\pm=1/2$, which gives Eq.~(\ref{SingleFluct_R}). Similarly,
for more general protocols, we have to analyze phase factors $\langle
\exp(i\int^t dt' D \, g(t') \, v_n(t'))\rangle$ with appropriate time
dependence of $g(t)$. In this case the first terms on the right
hand side of
Eqs.~(\ref{Eq:chi-}) are modified accordingly. For
the echo experiment we obtain in this way Eq.~(\ref{SingleFluct_E}).

The decay produced by a number of fluctuators is the
product of the individual contributions, i.e.,
$f_{R}(t)=\Pi_{n}\,f_{R,n}(t)$ and
$f_{E}(t)=\Pi_{n}\,f_{E,n}(t)$. If the noise is dominated by a
few fluctuators (this includes the case of many fluctuators in total,
but a few of them with similar rates $\gamma$) the fluctuations of
$X(t)$ may be strongly non-Gaussian.

\section{Non-Gaussian effects, $\mu<2$}

Since we consider uncorrelated TLF's the total decay of coherence
is the product of all single-TLF contributions, $f(t) = \prod_n
f_n(t)$, where $f_n(t)$ is given by \eqref{SingleFluct_R} and
\eqref{SingleFluct_E} for the free induction and echo experiment,
respectively. In Ref.~\cite{Galperin} an ensemble-averaged value of
$\ln f(t)$, denoted as $\langle \ln f(t) \rangle_F$, was calculated
for $\mu=1$. Here $\langle \cdots \rangle_F$ denotes the average over
the distribution of coupling strengths and switching rates
\eqref{eq:Pvgamma}. Both free induction decay and the ``phase memory decay''
(a protocol similar but not equivalent to the spin echo decay) were analyzed
in the regimes $t< \gamma_{max}^{-1}$ and $t> \gamma_{max}^{-1}$.
Below we will generalize these results to
the range $0<\mu<2$.

As discussed above, the quantity $\langle \ln f(t) \rangle_F$
is relevant for experiments with specific samples only if the
sample-to-sample fluctuations of $\ln f(t)$ are weak, i.e., if
$\ln f(t)$ is self-averaging. Then, the experimentally observable decay law $f(t)$
would be well approximated by $\exp(\langle \ln f(t) \rangle_F)$.
In Ref.~\cite{Galperin} the self-averaging was numerically
confirmed for the phase memory decay in the regime
$t< \gamma_{max}^{-1}$. Here we analyze the self-averaging in four regimes:
for the free induction and the echo cases, both in the limits
$t< \gamma_{max}^{-1}$ and $t> \gamma_{max}^{-1}$.
Specifically, we evaluate the ensemble-average $\langle
\ln f(t) \rangle_F$, given by an integral over the
$(v,\gamma)$-space. In some cases this integral is dominated by a range in the
`bulk' of the distribution, which contains many fluctuators on average;
this indicates that sample-to-sample fluctuations are weak. In other cases,
the integral is dominated by the boundary of the distribution, indicating that
the studied quantity is not self-averaging. Our analysis confirms the conclusion
of Ref.~\cite{Galperin}, obtained in one regime: for the echo decay at short times
$t< \gamma_{max}^{-1}$. We show further that in all other three regimes
investigated the dephasing law is not self-averaging.
In the calculations we assume that
$v_{\rm min}$, $\gamma_{\rm min}$ are very low frequency scales, and $1/t$
always exceeds them.
\subsection{Free induction decay}

For short times, $t< \gamma_{max}^{-1}$, we are effectively in the
static regime, and the ensemble-averaged free induction
decay is described by
\begin{equation}\label{lawFIDshort}
\langle \ln|f_{R}(t)| \rangle_F\propto -( D_{\lambda,z}\,\eta t)^\mu\ .%
\end{equation}
This result is dominated by the fluctuators with strength of order
$v\sim v_{\rm max}^{\rm typ}\sim\eta$ and thus is not
self-averaging. For an experiment with a specific sample the results
should be fitted by a contribution of one~(\ref{SingleFluct_R}) or a
few fluctuators, rather than by the ensemble-averaged
behavior~(\ref{lawFIDshort}). We can also estimate the typical decay
law for short times $t < \eta^{-1}$. In every realization there will
be a few strongest fluctuators, typically with strength $v_{\rm max}$. For $t\ll
v_{\rm max}^{-1}$ we obtain $\ln|f_{R}(t)|\approx - D_{\lambda,z}^2
t^2\,\sum_n v_n^2$. For distributions with $\mu < 2$ the sum $\sum_n
v_n^2$ is dominated by the largest $v_n$'s, and thus,
$\ln|f_{R}(t)|\propto - D_{\lambda,z}^2 t^2 v_{\rm max}^2$. Finally,
we can calculate the distribution function for the strength of the
strongest fluctuator $v_{\rm max}$ and obtain
\begin{equation}\label{distvmax}
P(v_{\rm max}) = \frac{\mu \eta^\mu}{v_{\rm max}^{1+\mu}}\,
e^{-(\eta/v_{\rm max})^\mu}\ .
\end{equation}
Most of the weight of this distribution is around $v_{\rm max}\sim
\eta$. Thus, in a typical sample for $t < \eta^{-1}$ the decay is given by
$\ln|f_{R}(t)|\propto - (D_{\lambda,z}\eta t)^2$ rather than by
(\ref{lawFIDshort}). To understand the difference we note that
the average decay law~(\ref{lawFIDshort}) can also be
obtained by averaging  the realization-dependent $-D_{\lambda,z}^2 v_{\rm max}^2t^2$
(valid for $t<v_{\rm max}^{-1}$) over the distribution (\ref{distvmax}). This
average is dominated by rare samples with a fluctuator of strength $v_{\rm max}\sim 1/t$
rather than by typical samples.

For longer times, $t > \gamma_{max}^{-1}$, the integration gives
\begin{itemize}
\item for $1\le\mu <2$
\begin{equation}
\langle \ln|f_{R}(t)|\rangle_F\propto -
\abs{\frac{\ln \gamma_{\rm min}t}
{\ln(\gamma_{\rm max}/\gamma_{\rm min})}}\,( D_{\lambda,z}\eta t)^\mu \ .\label{eq:Rdin1}%
\end{equation}

\item for $\mu<1$
\begin{equation}
\langle \ln|f_{R}(t)|\rangle_F \propto - c (D_{\lambda,z}\eta/\gamma_{\rm
max})^\mu  \gamma_{\rm max} t \ .\label{eq:Rdin2}%
\end{equation}
\end{itemize}

Both results are not self-averaging.

\subsection{Echo signal decay}

For short times, $t< \gamma_{max}^{-1}$, we find
\begin{equation}
\langle\ln|f_{E}(t)|\rangle_F \propto -c\, (D_{\lambda,z}\eta)^\mu \gmax
t^{1+\mu}\ ,
\end{equation}
For $c^{1/\mu}D_{\lambda,z} \eta > \gmax$ the echo decay is
dominated by this quasi-static contribution; the decay takes place
on the time scale shorter than the flip time of the fastest
fluctuators, $1/\gamma_{max}$. In this regime ($c^{1/\mu}
D_{\lambda,z}\eta > \gmax$) the result {\it is} self averaging since
it is dominated by fluctuators with $D_{\lambda,z} v \sim (c
D_{\lambda,z}^\mu \eta^\mu \gmax)^{1/(1+\mu)} < c^{1/\mu}
D_{\lambda,z} \eta < D_{\lambda,z} v_{\rm max}^{\rm typ}$.

For longer times, $t > \gamma_{max}^{-1}$ the dephasing is due to
multiple flips of the fluctuators. These times are
relevant if $c^{1/\mu} D_{\lambda,z}\eta < \gmax$.
The decay law is given by
\begin{itemize}
\item for $1<\mu<2$
\begin{equation}
\langle \ln|f_{E}(t)|\rangle_F\propto - c\, (D_{\lambda,z}\eta t)^\mu\ ,
\end{equation}
\item for $\mu=1$
\begin{equation}
\langle \ln|f_{E}(t)|\rangle F \propto - c\, (D_{\lambda,z}\eta t)
\ln(\gamma_{\rm max} t)\ ,
\end{equation}
\item for $\mu<1$
\begin{equation}
\langle \ln|f_{E}(t)|\rangle_F\propto - c (D_{\lambda,z}\eta/\gamma_{\rm
max})^\mu  \gamma_{\rm max} t \ .
\end{equation}
All these results are not
self-averaging.

\end{itemize}

\section{Quadratic coupling}

At the optimal working point, the first-order longitudinal coupling
$D_{\lambda,z}$ vanishes. Thus, to first order, the decay of the
coherent oscillations is determined by the relaxation processes and
for regular power spectra at low frequencies
one expects from Eq.~(\ref{Gamma2}) that $\Gamma_{2}=\Gamma_{1}/2$.
On the other hand, for power spectra which are singular at low frequencies
the second-order contribution of the longitudinal noise can be
comparable or even dominate over $\Gamma_{1}/2$. To evaluate this
contribution, one has to calculate
\begin{equation}
f_{2}(t)=\left\langle \exp\left(  i\,\frac{1}{2}\,\frac{\partial^{2}%
\omega_{01}}{\partial\lambda^{2}}\,\int\limits_{0}^{t} g(t')\,
X^{2}(t')d t'\right)  \right\rangle \,,\label{Eq:P}%
\end{equation}
where for the analysis of the free
induction decay (Ramsey signal) one sets $g(t')=1$, while for decay of
the echo-signal $g(t'<t/2)=-1$ and
$g(t'>t/2)=1$.

\textit{$1/f$ noise:} The free induction decay for the $1/f$ noise
with a high cutoff $\omega_{c}$ (the highest energy scale in the
problem) has been analyzed in Ref.~\cite{MakhlinPRL_X2}. Depending on
the time $t$, the decay is dominated by low- or high-frequency noise,
and the decay law can be approximated by a product of low-frequency
($\omega <1/t$, quasi-static) and high-frequency
($\omega>1/t$) contributions,
$f_{2,R}(t)=f_{2,R}^{\mathrm{lf}}(t)\cdot
f_{2,R}^{\mathrm{hf}}(t)$. The contribution of low frequencies is
given by~\cite{MakhlinPRL_X2,Rabenstein,Falci_Initial}
\begin{equation}
f_{2,R}^{\mathrm{lf}}(t)=\frac{1}{\sqrt{1-i\,\frac{\partial^{2}\omega_{01}%
}{\partial\lambda^{2}}\,\sigma_{X}^{2}t\,}}\,.\label{ramsey2}%
\end{equation}
For $1/f$ noise the variance of the low-frequency fluctuations is
$\sigma _{X}^{2}=2A\ln(1/\omega_{\mathrm{ir}}t)$. This
contribution  dominates at short times $t<\,\left[  (\partial^{2}\omega_{01}%
/\partial\lambda^{2})\,A/2\right]  ^{-1}$. At longer times, the
high-frequency contribution
\begin{equation}
\ln f_{2, R}^{\mathrm{hf}}(t)\approx-t\int\limits_{\sim1/t}^{\infty
}\frac{d\omega}{2\pi}\,\ln\left(  1-2\pi i\,\frac{\partial^{2}\omega_{01}%
}{\partial\lambda^{2}}\,S_{X}(\omega)\right)  \,,
\end{equation}
takes over. When $t\gg\left[
(\partial^{2}\omega_{01}/\partial\lambda
^{2})\,A/2\right]  ^{-1} $ (provided
$\omega_{c}\gg\pi\,(\partial^{2}\omega_{01}/\partial\lambda^{2})\,A
$) we obtain asymptotically $\ln f_{2, R}^{\mathrm{hf}%
}(t)\approx-(\pi/2)(\partial^{2}\omega_{01}/\partial\lambda^{2})\,At$.
Otherwise the quasistatic result (\ref{ramsey2}) is valid at all
relevant times. One can also evaluate the pre-exponential factor in
the long-time decay. This pre-exponent decays very slowly
(algebraically) but differs from 1 and thus further reduces
$f_{2,R}(t)$~\cite{JosefThesis}.

\textit{Quasi-static case:} In this case, i.e., when the cutoff
$\omega_{c}$ is lower than $1/t$ for all relevant times, the Ramsey
decay is simply given by the static contribution (\ref{ramsey2}). At
all relevant times the decay is algebraic and the crossover to the
exponential law is not observed. More generally, in the static
approximation with a distribution $P(\delta \lambda)$, the dephasing
law is given by the Fresnel-type integral transform,
\begin{equation}
f_{2,R}^{\mathrm{st}}(t)=\int d\,({X})\,P({X})\, \exp\left(i\,\frac{1}%
{2}\,\frac{\partial^{2}\omega_{01}}{\partial\lambda^{2}}\,{X}
^{2}\,t\right)\,,\label{ramsey_static_quad_PX}%
\end{equation}
which reduces to Eq.~(\ref{ramsey2}) for a Gaussian $P({X}
)\propto\exp\left(  -{X}^{2}/2\sigma_{X}^{2}\right)  $. In
general, any distribution $P({X})$, finite at ${X}=0$, yields a
long time decay of $f_{2,R}^{\mathrm{st}}$ proportional to
$t^{-1/2}$.

For $\mu < 2$ the analysis is technically more complicated. In that
case the distribution of initial conditions $P(X_0)$, and
equivalently the sum $X_0=\sum_{n}v_{n}\sigma_{n,z}(t=0)$, are no
longer Gaussian-distributed and, in particular, they cannot be
characterized by a typical width $\sigma$, due to the divergence of
the second moment $\expv{v^2}$ of~\eqref{eq:Pvgamma}. The
generalized central limit theorem tells us that $x=X_0/\eta$ is then
distributed according to a L\'evy distribution $L_{\mu,0}(x)$ and
consequently, according to \eqref{ramsey_static_quad_PX} the free
induction decay in the quasi-static regime is given by
\begin{equation}
\label{eq:PstmanyBF} f_{2,R}^{\mathrm{st}}(t) = \int dx \,
L_{\mu,0}(x) \;\exp\left(\frac{i}{2}\;
\frac{\partial^{2}\omega_{01}}{\partial\lambda^{2}}\, \eta^2 t
\cdot x^2\right)\,.
\end{equation}
For some values of $\mu$ explicit expressions are
known. An example is the Cauchy distribution, $L_{1,0}(x) =
1/[\pi(1+x^2)]$. Using \eqref{eq:PstmanyBF} the free induction decay
in the static regime, $t< \gmax^{-1}$, is then given by
\begin{equation}
  \label{eq:PstmanyBFmu1}
   f_{2,R}^{\mathrm{st}}(t) = e^{-i\alpha t}
   \left[1 - \Phi\left(\sqrt{\alpha t/i}\right)\right] \,.
\end{equation}
Here we introduced the rate
\begin{equation}
\label{eq:Defalphamu1} \alpha = \frac{1}{2}
\frac{\partial^{2}\omega_{01}}{\partial\lambda^{2}}\,\eta^2
\end{equation}
and $\Phi(z) = 2\pi^{-1/2} \int_0^z dx\, e^{-x^2}$ denotes the
error function. One can expand $\Phi(z)$ in
\eqref{eq:PstmanyBFmu1} to find the asymptotic behavior of
$f_{2,R}^{\mathrm{st}}(t)$ for $\mu=1$:
\begin{equation}
  \label{eq:PstmanyBFmu1a}
  \abs{f_{2,R}^{\mathrm{st}}(t)} = \left\{
    \begin{array}{l @{\quad {\rm for} \quad}l}
    1 - \left(\frac{2}{\pi}\,\alpha t\right)^{1/2} & t \ll \alpha^{-1} \vspace*{0.2cm}\\
\left(\pi \alpha t\right)^{-1/2} & t \gg \alpha^{-1}
    \end{array}\right.\;.
\end{equation}
The initial decay for $t \ll \alpha^{-1}$ is thus very fast, but at
times $t \approx \alpha^{-1}$ the decay crosses over to a much
slower power law $\propto 1/\sqrt{t}$. The dephasing time scales as
$\alpha^{-1}$, but with a relatively large prefactor
due to the slow algebraic decay. For other values of $\mu<2$ the
asymptotic behavior of $f_{2,R}^{\mathrm{st}}$ has been obtained in
Ref.~\cite{JosefThesis}:
\begin{equation}
  \label{eq:PstmanyBFmu1b}
  \abs{f_{2,R}^{\mathrm{st}}(t)} = \left\{
    \begin{array}{l @{\quad {\rm for} \quad}l}
    1 - C(\mu)\left(\alpha t\right)^{\mu/2} &  t \ll \alpha^{-1} \vspace*{0.2cm}\\
    D(\mu)\left(\alpha t\right)^{-1/2} &  t \gg \alpha^{-1}
    \end{array}\right.\;,
\end{equation}
where $C(\mu)$ and $D(\mu)$ are factors of order 1.

Let us now discuss the shape of the decay of
$f_{R}^{\mathrm{st}}$ qualitatively and comment on their
validity. The initial decay of $f_{R}^{\mathrm{st}}$ for $\mu<2$
is singular and thus very fast compared to the Gaussian case
$\mu>2$. This initial decay is dominated by strongly coupled
fluctuators, i.e., by the tail of the distribution
(\ref{eq:Pvgamma}). It is, thus, not self-averaging.

On the other hand, for longer times, $t \gg \alpha^{-1}$, the
decay goes over to a much slower power law. The exponent $-1/2$ is
independent of $\mu$ and coincides even with the
prediction of the Gaussian model $(\mu>2)$. Hence, the
$1/\sqrt{t}$ decay law appears to be universal in the presence of
quasi-static noise, independent of the considered statistics.
For low enough $\gmax$ such that $\alpha \gg \gmax$ the free
induction signal decays already in the quasi-static regime, $t <
\gmax^{-1}$, and is thus given by \eqref{eq:PstmanyBFmu1b}.
Otherwise, further analysis characterizing the contribution of
fast fluctuators, $\gamma > t^{-1}$, is needed to describe
decoherence.

\section{Conclusions}

We have shown that non-Gaussian $1/f$ noise of
ensembles of two-level fluctuators frequently leads to
non-self-averaging dephasing laws. Non-Gaussian noise
arises, for instance, when the distribution of coupling
strengths between the two-level fluctuators and a qubit has a long
algebraic tail. In this case, since experiments are
performed on specific samples,
one should study the typical rather than ensemble averaged
behavior. Interestingly, in certain regimes, e.g., for short-time
echo decay, the decay law is self-averaging.

\section{Acknowledgement}

The work is part of the CFN of the DFG and of the
EU IST Project SQUBIT.
YM acknowledges support from the Dynasty Foundation
and the grant MD-2177.2005.2.

\vskip 1cm

\bibliographystyle{unsrt}
\bibliography{ref}

\begin{thebibliography}{10}

\bibitem{Nakamura_Echo}
Y.~Nakamura, \mbox{Yu.} A.~Pashkin, T.~Yamamoto, and J.~S. Tsai.
\newblock Charge echo in a {C}ooper-pair box.
\newblock {\em Phys. Rev. Lett.}, 88:047901, 2002.

\bibitem{Saclay_Science}
D.~Vion, A.~Aassime, A.~Cottet, P.~Joyez, H.~Pothier, C.~Urbina, D.~Esteve, and
  M.~H. Devoret.
\newblock Manipulating the quantum state of an electrical circuit.
\newblock {\em Science}, 296:886, 2002.

\bibitem{Clarke}
D.~J. VanHarlingen, T.~L. Robertson, B.~L.~T. Plourde, P.~A. Reichardt, T.~A.
  Crane, and John Clarke.
\newblock Decoherence in {J}osephson-junction qubits due to critical current
  fluctuations.
\newblock {\em Phys. Rev. B}, 70:064517, 2004.

\bibitem{Aguado}
R.~Aguado and L.~P. Kouwenhoven.
\newblock Double quantum dots as detectors of high-frequency quantum noise in
  mesoscopic conductors.
\newblock {\em Phys. Rev. Lett.}, 84:1986, 2000.

\bibitem{SchoelkopfNoiseDetector}
R.~J. Schoelkopf, A.~A. Clerk, S.~M. Girvin, K.~W. Lehnert, and M.~H. Devoret.
\newblock Qubits as spectrometers of quantum noise.
\newblock In Yuli~V. Nazarov, editor, {\em Quantum Noise in Mesoscopic
  Physics}, pages 175--203, Dordrecht, Boston, 2003. Kluwer Academic
  Publishers.
\newblock cond-mat/0210247.

\bibitem{Astafiev}
O.~Astafiev, \mbox{Yu}. A.~Pashkin, Y.~Nakamura, T.~Yamamoto, and J.~S. Tsai.
\newblock Quantum noise in the {J}osephson charge qubit.
\newblock {\em Phys. Rev. Lett.}, 93:267007, 2004.

\bibitem{Ithier2005}
G.~Ithier, E.~Collin, P.~Joyez, P.~J. Meeson, D.~Vion, D.~Esteve, F.~Chiarello,
  A.~Shnirman, {Yu.} Makhlin, J.~Schriefl, and G.~Sch{\"o}n.
\newblock Decoherence in a superconducting quantum bit circuit.
\newblock {\em Phys. Rev. B}, 72:134519, 2005.

\bibitem{Our_PhysScripta}
A.~Shnirman, {\relax Yu}.~Makhlin, and G.~Sch{\"o}n.
\newblock Noise and decoherence in quantum two-level systems.
\newblock {\em Physica Scripta}, T102:147, 2002.

\bibitem{Paladino_PRL02}
E.~Paladino, L.~Faoro, G.~Falci, and R.~Fazio.
\newblock Decoherence and 1/f noise in {J}osephson qubits.
\newblock {\em Phys. Rev. Lett.}, 88:228304, 2002.

\bibitem{Galperin}
Y.~M. Galperin, B.~L. Altshuler, and D.~V. Shantsev.
\newblock Low-frequency noise as a source of dephasing of a qubit.
\newblock In I.~V. Lerner, B.~L. Altshuler, and {Yu.} Gefen, editors, {\em
  Fundamental Problems of Mesoscopic Physics}, Dordrecht, Boston, London, 2004.
  Kluwer Academic Publishers.
\newblock cond-mat/0312490.

\bibitem{Bernamont}
J.~Bernamont.
\newblock {\em Ann. Phys. (Leipzig)}, 7:71, 1937.

\bibitem{Ludviksson}
A.~Ludviksson, R.~Kree, and A.~Schmid.
\newblock Low-frequency $1/f$ fluctuations of resistivity in disordered metals.
\newblock {\em Phys. Rev. Lett.}, 52:950, 1984.

\bibitem{Kogan}
Sh.~M. Kogan and K.~E. Nagaev.
\newblock On the low-frequency current noise in metals.
\newblock {\em Solid State Comm.}, 49:387, 1984.

\bibitem{Kozub}
V.~I. Kozub.
\newblock Properties of superconducting {S-I-N}, {S-I-S}, and {S-C-S}
  structures with amorphous weak coupling.
\newblock {\em Sov. Phys. JETP}, 60:810, 1984.

\bibitem{Galperin_Gurevich_Kozub}
Y.~M. Galperin, V.~L. Gurevich, and V.~I. Kozub.
\newblock Disorder-induced low-frequency noise in small systems - point and
  tunnel contacts in the normal and superconducting state.
\newblock {\em Europhys. Lett.}, 10:753, 1989.

\bibitem{Stamp}
N.~V. Prokof'ev and P.~C.~E. Stamp.
\newblock Theory of the spin bath.
\newblock {\em Rep. Prog. Phys.}, 63:669, 2000.

\bibitem{DuttaHorn}
P.~Dutta and P.~M. Horn.
\newblock Low-frequency fluctuations in solids: $1/f$ noise.
\newblock {\em Rev. Mod. Phys.}, 53:497, 1981.

\bibitem{Bloch_Derivation}
F.~Bloch.
\newblock Generalized theory of relaxation.
\newblock {\em Phys. Rev.}, 105:1206, 1957.

\bibitem{Redfield_Derivation}
A.~G. Redfield.
\newblock On the theory of relaxation processes.
\newblock {\em IBM J. Res. Dev.}, 1:19, 1957.

\bibitem{ACthesis}
A.~Cottet.
\newblock Implementation of a quantum bit in a superconducting circuit.
\newblock {\em PhD thesis}, Universit\'{e} Paris VI, 2002.

\bibitem{MakhlinPRL_X2}
Yu. Makhlin and A.~Shnirman.
\newblock Dephasing of solid-state qubits at optimal points.
\newblock {\em Phys. Rev. Lett.}, 92:107001, 2004.

\bibitem{Rabenstein}
K.~Rabenstein, V.~A. Sverdlov, and D.~V. Averin.
\newblock Qubit decoherence by gaussian low-frequency noise.
\newblock {\em JETP Lett.}, 79:783, 2004.

\bibitem{Falci_Initial}
G.~Falci, A.~D'Arrigo, A.~Mastellone, and E.~Paladino.
\newblock Initial decoherence in solid state qubits.
\newblock {\em Phys. Rev. Lett.}, 94:167002, 2005.

\bibitem{JosefThesis}
J.~Schriefl.
\newblock Dekoh{\"a}renz in {J}osephson quantenbits.
\newblock {\em PhD Thesis}, University Karlsruhe, 2005.

\end{thebibliography}

\end{document}